\documentclass[useAMS,usenatbib]{mn2e}

\usepackage[pdftex]{graphicx}
\usepackage{epstopdf}
\usepackage{graphicx}
\usepackage[fleqn]{amsmath}

\newcommand{\rp}{r_{\mathrm p}}

\newcommand{\op}{\Omega_\mathrm{p}}

\newcommand{\vkep}{v_\mathrm{K}}
\newcommand{\cs}{c_\mathrm{s}}

\newcommand{\xs}{x_\mathrm{s}}

\newcommand{\be}{ \begin {equation}}
\newcommand{\ee}{ \end {equation}}

\newcommand{\sigp}{\Sigma_\mathrm{p}}

\title[Dynamical corotation torques]{Dynamical corotation torques on low-mass planets}
\author[S.-J. Paardekooper]{S.-J. Paardekooper$^{1, 2}$\thanks{E-mail:
s.j.paardekooper@qmul.ac.uk}\\
$^1$Astronomy Unit, School of Physics and Astronomy, Queen Mary, University of London, \\ \indent Mile End Road, London E1 4NS, United Kingdom\\
$^2$DAMTP, University of Cambridge, Wilberforce Road, Cambridge CB3 0WA,
United Kingdom}

\begin{document}

\date{Draft version \today}

\pagerange{\pageref{firstpage}--\pageref{lastpage}} \pubyear{2011}

\maketitle

\label{firstpage}

\begin{abstract}
We study torques on migrating low-mass planets in locally isothermal discs. Previous work on low-mass planets generally kept the planet on a fixed orbit, after which the torque on the planet was measured. In addition to these static torques, when the planet is allowed to migrate it experiences dynamical torques, which are proportional to the migration rate and whose sign depends on the background vortensity gradient. We show that in discs a few times more massive than the Minimum Mass Solar Nebula, these dynamical torques can have a profound impact on planet migration. Inward migration can be slowed down significantly, and if static torques lead to outward migration, dynamical torques can take over, taking the planet beyond zero-torque lines set by saturation of the corotation torque in a runaway fashion. This means the region in non-isothermal discs where outward migration is possible can be larger than what would be concluded from static torques alone.
\end{abstract}


\section{Introduction}

Ever since \cite{goldreich80} it has been known that satellites embedded in Keplerian discs are very mobile. Based on theoretical arguments, significant orbital migration of newly formed planets should be the rule rather than the exception. There are strong indications in the population of extrasolar planets \citep{baruteau13}, and perhaps even the Solar system \citep{walsh11}, that disc migration indeed has played a role in shaping planetary systems.    

Disc migration traditionally comes in three flavours. Low-mass planets, roughly up to the mass of Neptune, migrate through the excitation of linear density waves in the disc, plus a contribution from the corotation region. This is called Type I migration. Early analytical \citep{tanaka02} and numerical work \citep[e.g.][]{nelson00} focused on isothermal discs, in which the temperature is prescribed and fixed. It was found that migration is always directed inward for reasonable disc parameters, and that migration time scales are much shorter than the disc life time, so that according to migration theory \emph{all} planets should end up very close to the central star. This is of course not what is observed, and therefore a lot of work has gone into finding mechanisms that limit the range of inward migration. 

While Type I migration was always thought to be due to \emph{linear} interactions with the disc, it was shown in \cite{drag} that the corotation torque in isothermal discs is in fact nonlinear, and that this nonlinear corotation torque or horseshoe drag \citep{ward91} can be much larger than previous linear estimates, working against fast inward migration. Nevertheless, it took a departure from the isothermal assumption \citep{paard06} to really see the potential of nonlinear corotation torques and even migration reversal. Much work has been going into understanding the nature of these torques and under what circumstances migration can be outward \citep{masset09, paardekooper10, masset10, paardekooper11}. A recent overview can be found in \cite{baruteau13}. 

A major drawback of relying on corotation torques to halt or even reverse hazardous inward migration, is that they are prone to saturation. In the absence of any diffusive process, the corotation region is a closed system, and can therefore only provide a finite amount of angular momentum to the planet. Usually viscosity is invoked to keep the torque from saturating \citep{masset01,masset10, paardekooper11}, which makes the migration rate depend strongly on the location in the disc \citep{bitsch13}. The limiting radius to which a planet can migrate outward in radiative discs appears to be set by saturation of the corotation torque \citep{bitsch11}. This means that as planets grow more massive, it becomes harder to keep them at large orbital radii. 

Higher-mass planets, comparable in mass to Jupiter, carve deep gaps in the disc \citep[e.g.][]{lin86a}. While the gap inhibits close-range disc-planet interactions responsible for Type I migration, the planet is locked in the gap and will migrate with the rest of the disc at the local viscous accretion rate \citep{lin86b, crida07} if the local disc mass is much larger than the planet mass, or at a reduced rate if the planet is more massive \citep{syer95}. This mode of migration is called Type II. In this work, we will be dealing exclusively with non-gap opening planets.

A third flavour of migration, Type III, was originally called runaway migration \citep{masset03}. This because is relies on coorbital torques whose magnitude is proportional to the migration rate. Depending on the sign of this dynamical torque (dynamical because it needs the planet to be migrating in order to operate), there is the possibility of a positive feedback loop \citep{masset08}. This sign is determined by the \emph{coorbital mass deficit}: if the coorbital region is depleted in mass, for example if the planet carves a partial gap in the disc, the dynamical torque reinforces the migration rate, leading to fast Type III migration if the mass deficit is comparable to the planet mass. 

Since these dynamical torques reinforce an existing migration rate, Type III migration can be directed both inwards and outwards \citep{masset03}. If outward migration can be set up, together with a positive coorbital mass deficit, outward Type III migration will result. However, because of accretion onto the planet it appears that the scope of outward Type III migration is limited, at least for planets comparable in mass to Jupiter  \citep{peplinski08c}. 

Type III migration was originally associated with intermediate-mass planets, in between the Type I and Type II migration regime, since these planets carve partial gaps and hence have a coorbital mass deficit. More massive planets need an extremely massive disc, beyond the gravitational stability limit, for the mass deficit to be comparable to the planet mass \citep{masset03}. However, it is possible to rely on the disc structure itself, rather than the planet modifying it, to set up a coorbital mass deficit. If a planet finds itself on a steep density gradient, a mass deficit naturally occurs, and Type III migration can be initiated even for Jupiter-like planets \citep{peplinski08}. In principle, one could imagine a similar situation arising for low-mass planets. However, we will see that, in the case of low-mass planets, it is not the density gradient that matters, but rather the \emph{vortensity}\footnote{Vortensity is defined as the ratio of vorticity and surface density} gradient. Such a gradient can give rise to dynamical torques reminiscent of Type III migration. 

Dynamical corotation torques can play a role for low-mass planets, especially in the regime where the static corotation torques (i.e. torques independent of the migration rate) saturate. This is the subject of this paper. The plan of this paper s as follows. In section \ref{secEq}, we introduce the disc models used, and in section \ref{secNum} we describe our numerical method. In section \ref{secMig} we present a simple model for migration rates including dynamical torques, and check the prediction of this model with numerical simulations in section ref{secRes}. The results are discussed in section \ref{secDisc} and we conclude in section \ref{secCon}.
 
\section{Basic equations and disc models}
\label{secEq}

We will work in the two-dimensional (2D) approximation (in cylindrical polar coordinates $(r,\varphi)$, centred on the central star), where all quantities are treated as vertical averages over the vertical extent of the disc, which is assumed to be isothermal with a fixed radial temperature gradient (i.e. locally isothermal). The basic equations are then mass conservation and Euler's equation:
\begin{eqnarray}
\frac{\partial\Sigma}{\partial t} + \nabla \cdot (\Sigma{\bf v})&=&0\\
\frac{\partial{\bf v}}{\partial t} +{\bf v} \cdot\nabla {\bf v} &=& -\frac{\nabla p}{\Sigma}-\nabla \Phi + {\bf f}_\nu,
\label{eqEuler}
\end{eqnarray}
where $\Sigma$ is the surface density and ${\bf v}$ the velocity. The pressure $p = \cs^2\Sigma$, where the sound speed $\cs$ has a radial profile such that the aspect ratio of the disc $h=H/r=\cs/\vkep$ is constant, with $H$ the scale height of the disc and $\vkep$ the Keplerian velocity. The last term in equation (\ref{eqEuler}) represents viscous forces, the exact form of which can be found for example in \cite{dangelo02}.

The gravitational potential $\Phi$ is made up of three components (we do not consider the self-gravity of the disc): gravity from the central star, gravity due to the planet, and an indirect term due to the acceleration of the coordinate frame:
\begin{equation}
\Phi = -\frac{GM_*}{r}-\frac{GM_p}{|{\bf r} - {\bf r}_\mathrm{p}|}+\frac{GM_p}{\rp^3}{\bf r}\cdot{\bf r}_\mathrm{p},
\end{equation}
where subscripts p indicate quantities related to the planet. The second term needs to be softened in 2D, with a softening length $\epsilon$ that should be comparable to $H$. In this paper, we use $\epsilon=0.4H$, which gives good agreement with 3D linear theory in the linear regime \citep{paardekooper10}. We will be concerned with planets that do not open up gaps in the disc, and therefore we will limit the mass ratio $q = M_p/M_* \ll h^3$ \citep{kory96}. Our standard planetary mass has $q=10^{-5}=0.08h^3$, which amounts to roughly $3$ $M_\oplus$ around a Solar mass star.   

For simplicity, we will work with discs that are power laws in surface density, with power law index $-\alpha$. We parametrise the mass of the disc through 
\begin{equation}
q_d=\frac{\pi r_0^2\Sigma_0}{M_*},
\label{eqQd}
\end{equation}
where $\Sigma_0$ is the surface density at reference radius $r_0$ (the initial orbital radius of the planet). For an overview of dimensionless quantities used in this work, see table \ref{tabDim}. The Minimum Mass Solar Nebula \citep[MMSN,][]{hayashi81} has $q_d=0.002$ at 10 AU. Note that the Toomre stability parameter $Q=h/q_d$. 

\begin{table}
\begin{tabular}{@{}l@{}l@{}l@{}c}
\hline
Symbol\,\,\,\,\, & Meaning\, & Unit & Equation\\
\hline
$q$ & planet mass & stellar mass\\
$q_d$ & disc mass & stellar mass & (\ref{eqQd})\\
$h$ & disc aspect ratio & & \\
$\gamma$ & torque & $\Gamma_0$ & (\ref{eqGamma0})\\
$\zeta$ & orbital radius & initial orbital radius &\\
$\tau$ & time & migration time scale & (\ref{eqTmig})\\
$m_c$ & coorbital mass & planet mass & (\ref{eqmc})\\
$\bar\xs$ & horseshoe width\,\,\,\, & orbital radius &\\
\hline
\end{tabular}
\caption{Overview of dimensionless quantities used in this work. For example, $q_d$ denotes the disc mass in terms of mass of the central star, and is defined in equation (\ref{eqQd}).}
\label{tabDim} 
\end{table}

We choose a radial dependence of kinematic viscosity $\nu$ such that the initial surface density profile is an equilibrium solution, which means we need
\begin{equation}
\nu(r) = \nu_0 \left(\frac{r}{r_0}\right)^{\alpha-1/2},
\label{eqnu}
\end{equation} 
for some reference radius $r_0$, which we take to be the initial orbital radius of the planet. This greatly simplifies the analysis, since there will be no accretion flow, and the disc does not viscously evolve away from the initial profile. 

\section{Numerical method}
\label{secNum}

We use the 2D hydrodynamics code {\sc FARGO} \citep[Fast Advection in Rotating Gaseous Objects,][]{masset00} in its isothermal version without self-gravity, which was adapted to include the viscosity law (\ref{eqnu}). When calculating the forces onto the planet, all disc material was taken into account. For low-mass planets, that do not have gravitationally bound dense envelopes, there is no need to exclude material inside their Hill radius. The axisymmetric component of the force was neglected to ensure consistency as the planet migrates \citep{baruteau08}.

The computational domain extends radially from $r_\mathrm{min}=0.4r_0$ to $r_\mathrm{max}=2.5r_0$ and the full $2\pi$ in azimuth. This domain is covered by a uniform grid with typically $n_r=512$ cells in the radial and $n_\varphi=1536$ cells in the azimuthal direction. For cases with outward migration, the radial size of the domain is doubled to $r_\mathrm{max}=4.6r_0$, while doubling the number of radial grid cells.
 
\section{Migration due to static and dynamical torques}
\label{secMig}

Below, we will distinguish between \emph{static} torques (i.e. torques independent of the migration rate) and \emph{dynamical} torques (i.e. torques that do depend on the migration rate). For Type I migration, it is the static torque is what is usually measured and modelled \citep[e.g.][]{tanaka02, masset09, masset10, paardekooper10, paardekooper11}. 

\subsection{Static torques}
In our locally isothermal disc, with constant $h$ and the unperturbed surface density a power law in radius, the planet will experience a torque $\Gamma$ from the disc \citep{goldreich80}. Assuming the orbit of the planet stays circular, the orbital radius evolves as
\begin{equation}
\frac{d\rp}{dt} = 2\rp \frac{\Gamma/\Gamma_0}{M_\mathrm{p} \sqrt{GM_*\rp}}\Gamma_0,
\label{eqdrpdt1}
\end{equation}
with 
\begin{equation}
\Gamma_0 = \frac{q^2}{h^2} \sigp \rp^2\op^2 = \frac{q^2}{h^2}\Sigma_0r_0^4\Omega_0^2\left(\frac{\rp}{r_0}\right)^{1-\alpha},
\label{eqGamma0}
\end{equation}
where subscripts $0$ indicate quantities evaluated at the \emph{initial} orbital radius of the planet. The quantity $|\Gamma|/\Gamma_0$ is usually of order unity \citep{paardekooper10}. Using the expression for $\Gamma_0$, equation (\ref{eqdrpdt1}) transforms into
\begin{equation}
\frac{1}{r_0\Omega_0}\frac{d\rp}{dt}=2\frac{\Gamma}{\Gamma_0}\frac{q}{h^2}\frac{\Sigma_0r_0^2}{M_*}\left(\frac{\rp}{r_0}\right)^{3/2-\alpha}.
\label{eqdrdt0}
\end{equation}
If $\Gamma/\Gamma_0$ does not depend on $\rp$ (or explicitly on $t$), this equation is easily integrated, for $\alpha \neq 1/2$, to
\begin{equation}
\frac{\rp(t)}{r_0}=\left[1+\frac{2}{\pi}\frac{\Gamma}{\Gamma_0}\frac{q_d q}{h^2}\left(\alpha-\frac{1}{2}\right)t\right]^{\frac{2}{2\alpha-1}},
\label{eqdrdt}
\end{equation}
with $q_d$ a measure of the disc mass (see equation (\ref{eqQd})), while for $\alpha=1/2$ simple exponential decay results\footnote{Note that in fact the limit $\alpha\rightarrow 1/2$ exists and gives exponential decay} . Note that $q_d < h$ for gravitationally stable discs. 

One can define a migration time scale
\begin{equation}
\tau_\mathrm{mig} = \frac{\pi}{2}\frac{h^2}{q_dq}\Omega_0^{-1},
\label{eqTmig}
\end{equation}
and use it to cast equation (\ref{eqdrdt0}) in dimensionless form:
\begin{equation}
\frac{d\zeta}{d\tau}=\gamma \zeta^{3/2-\alpha},
\end{equation}
where $\gamma = \Gamma/\Gamma_0$, $\zeta=\rp/r_0$ and $\tau = t/\tau_\mathrm{mig}$.

Note that the wave torque automatically satisfies the condition on $\gamma$ (to be independent of $\rp$ and $t$), as well as the unsaturated corotation torque. If the corotation torque partially saturates, its magnitude can vary with $\rp$ in a non-trivial way depending on how viscosity varies with radius. Saturation is governed by the ratio $p^2$ of the diffusion time scale across the horseshoe region and the libration time scale:
\begin{equation}
p^2=\frac{\tau_\nu}{\tau_\mathrm{lib}}=\frac{2}{3}\frac{\xs^3}{\nu}\frac{\op}{\rp}\propto \frac{1}{\nu}\sqrt{\rp},
\label{eqP}
\end{equation} 
where $\xs$ is the half-width of the horseshoe region, which can be estimated for low-mass planets as $\xs \approx \rp \sqrt{q/h}$ \citep{horse}. This means that for $\nu\propto \sqrt{r}$, saturation of the corotation torque is similar throughout the disc. Comparing this with the condition that the initial surface density profile should be an equilibrium solution, we find that this happens for $\alpha=1$. Therefore, pure power-law migration, as expressed in equation (\ref{eqdrdt}), occurs in isothermal discs for $\alpha =1$ and $\alpha=3/2$, since in the latter case there is no corotation torque and therefore no saturation to worry about. Note that in \emph{locally} isothermal discs, additional effects play a role, which we neglect here for simplicity \citep{casoli09}.

\begin{figure}
\centering
\resizebox{\hsize}{!}{\includegraphics[]{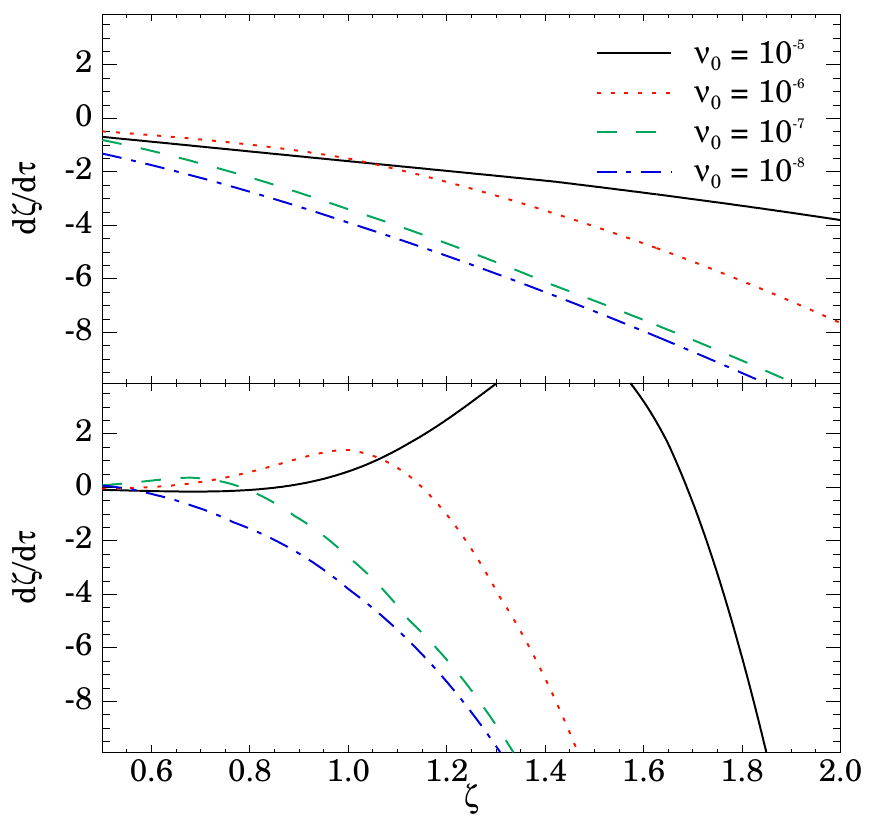}}
\caption{Migration rate versus radius for static torques for different levels of viscosity at $\zeta=1$. Top panel: $\alpha=0$, bottom panel: $\alpha = -2$.}
\label{figStatic}
\end{figure}

We have measured the static torque $\gamma$ as a function of viscosity, keeping a planet with $q=10^{-5}$ on a fixed circular orbit in a disc with $h=0.05$. Two values of $\alpha$ were considered: $\alpha=0$ and $\alpha=-2$. The latter case, while unrealistic for the disc as a whole, gives very strong corotation torques that can reverse the direction of migration, and can therefore serve as an approximation of what could happen in non-isothermal discs with strong entropy-related torques. In figure \ref{figStatic} we show the migration rates due to static torques as a function of radius. Towards larger values of $\zeta$, corotation torques saturate, and $d\zeta/d\tau$ approaches $\gamma_L \zeta^{3/2-\alpha}$.

\begin{figure}
\centering
\resizebox{\hsize}{!}{\includegraphics[]{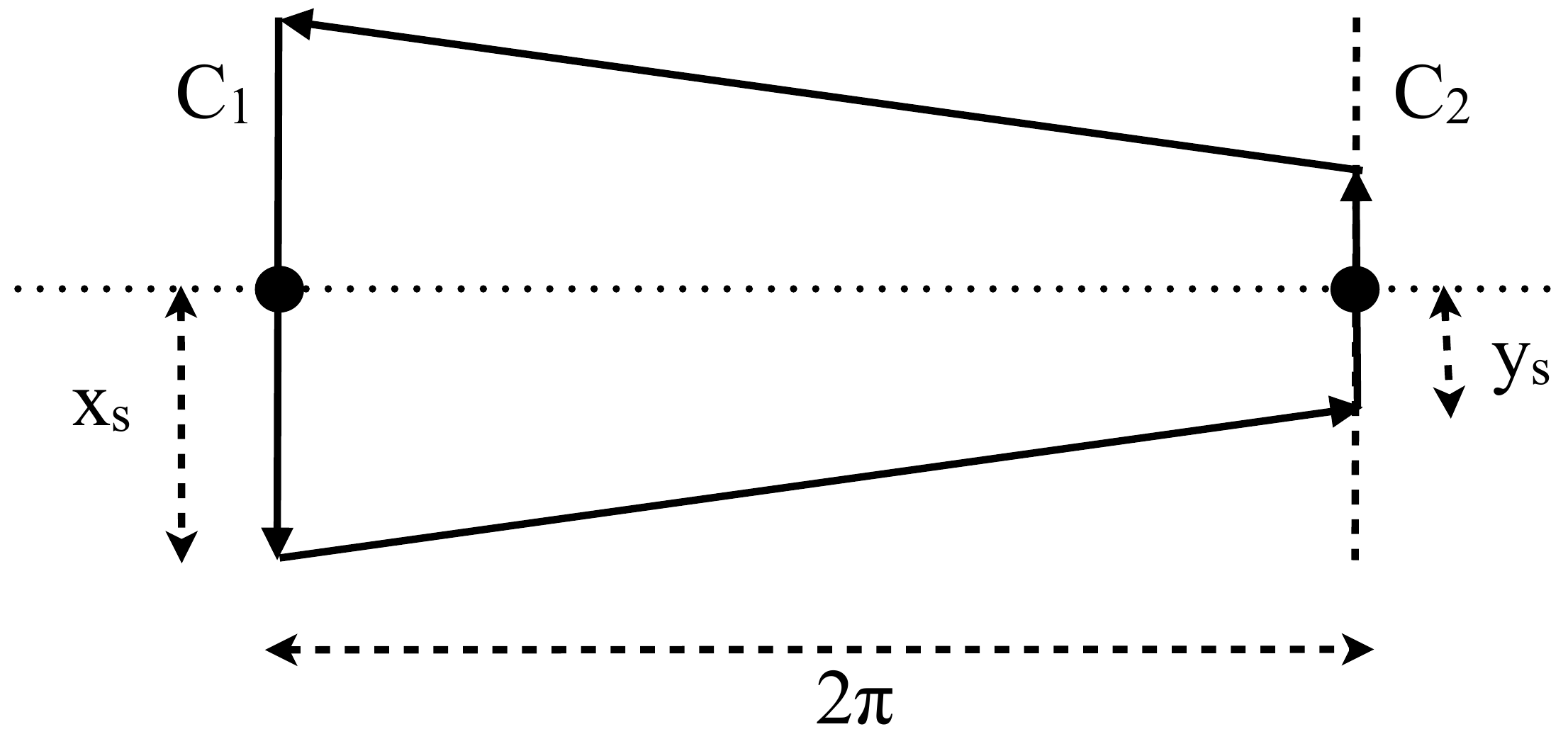}}
\caption{Schematic view of the separatrix (solid lines) in a coordinate frame moving with the planet, which is migrating inward towards the bottom of the figure. The vertical axis represents the radial distance to the planet; the horizontal axis represents azimuth. The planet is indicated by the black circles.}
\label{figSchematic}
\end{figure}

\subsection{Dynamical corotation torque}
The corotation torque on a migrating planet was worked out in detail in \cite{masset03}. A shortened version is presented here, applied to low-mass planets, highlighting the main assumptions.

For a migrating planet, the corotation region gets deformed. Between horseshoe turns, the gas drifts outward (inward) for an inward (outward) migrating planet. The shape of the horseshoe region is schematically depicted in figure \ref{figSchematic} for an inward migrating planet. Throughout this section, we will assume \emph{slow} migration in the sense that the radial drift of gas in between horseshoe turns is small compared to the unperturbed horseshoe width \citep[see also][]{masset03}:
\begin{equation}
\left| \frac{d\rp}{dt}\right| \ll \frac{3\op \xs^2}{4\pi\rp}.
\end{equation}
Physically, this means that the time it takes for the planet to migrate across its own horseshoe region is long compared to the libration time scale. In this case, on the left side of figure \ref{figSchematic}, the horseshoe turns have amplitude $\xs$, which we take to be the unperturbed value of the width of the horseshoe region.  This is a reasonable assumption in the case of slow migration. On the right side of figure \ref{figSchematic}, the horseshoe turns have lower amplitude $y_\mathrm{s}$, due to the outward radial drift of gas parcels relative to the planet:
\begin{equation}
y_\mathrm{s} = \xs + \frac{4\pi \rp}{3\op\xs}\frac{d\rp}{dt}.
\label{eqYs}
\end{equation}
Note that $d\rp/dt <0$ and therefore $y_\mathrm{s} < \xs$.

In general, the corotation torque can be represented as an integral over the inverse over the specific vorticity $w$, evaluated on two sides of the planet \citep{masset03,drag}:
\begin{equation}
\Gamma_\mathrm{hs}=\frac{3}{4} \rp\op^3 \int_0^{\xs} \left(w_{C_1}(x)-w_{C_2}(x)\right)x^2dx,
\label{eqTorqueInt}
\end{equation}
where $x = r-\rp$, and we have assumed a Keplerian rotation profile. The two lines of integration $C_1$ and $C_2$ are shown in figure \ref{figSchematic}. 

Now we make the assumption that $w$ inside the corotation region can be represented by a single value $w=w_c$. For example, if the corotation torque is saturated completely, material inside the horseshoe region (inside the separatrix depicted in figure \ref{figSchematic}) has been phase-mixed to an extent that it can be represented by a single value of $w=w_c$. Because the corotation region migrates with the planet, in an inviscid disc this value can be taken as the inverse of the specific vorticity at the initial location of the planet, $w_c=w(r_0)$. In the case of viscous discs, a different value of $w_c$ has to be adopted (see section \ref{secVisc} below). We furthermore neglect any variation in $w$ outside the corotation region, which is therefore represented by $w(\rp)$. Using this in equation (\ref{eqTorqueInt}), we get
\begin{equation}
\Gamma_\mathrm{hs}=\frac{3}{4} \rp\op^3 \int_{y_\mathrm{s}}^{\xs} \left(w_c-w(\rp)\right)x^2dx.
\end{equation} 
Assuming migration is slow compared to the libration time scale, we have $y_\mathrm{s} \approx \xs$, and above integral simplifies to
\begin{equation}
\Gamma_\mathrm{hs}=\frac{3}{4} \rp\op^3 (\xs-y_\mathrm{s}) \left(w_c-w(\rp)\right) \xs^2,
\end{equation}
which, with help of equation (\ref{eqYs}), becomes
\begin{equation}
\Gamma_\mathrm{hs} = 2\pi \left(1-\frac{w_c}{w(\rp)}\right)\sigp\rp^2x_s\op\frac{d\rp}{dt}.
\end{equation}  
The factor $1-w_c/w(\rp)$ plays the same role as the coorbital mass deficit in classical Type III migration \citep{masset03}. The mass deficit is usually positive for intermediate-mass planets because of the surface density depression around the orbit of the planet. The dynamical corotation torque then has a positive feedback on migration: if migration is inward, the dynamical torque is negative, leading to runaway behaviour \citep{masset03}. In the case of a low-mass planet, with very little surface density perturbations, it is rather a \emph{vortensity} deficit. This deficit arises purely because of migration, and can be of either sign depending on the background vortensity gradient in the disc. If the planet is migrating towards a region of higher vortensity (i.e. lower $w$), $w_c  = w(r_0) > w(\rp)$, and the vortensity deficit is negative. Note that this case corresponds to a positive static corotation torque. This means that if the planet is migrating in the opposite direction of what the corotation torque wants it to do, the dynamical torque has a negative feedback on migration. On the other hand, if the planet is migrating in the same direction as the static corotation torque wants it to go, there is a positive feedback with the possibility of runaway effects. Since, if corotation torques are completely saturated, the only other torque on the planet is the wave torque, which is usually negative, we will be dealing with inward migration only in inviscid discs. Depending on the background vortensity gradient, the dynamical torque exerts a negative feedback on migration ($\alpha < 3/2$), or a positive feedback ($\alpha > 3/2$).  
 
Orbital evolution is then due to the static torque and the dynamical corotation torque:
\begin{eqnarray}
\frac{d\rp}{dt}&=&\frac{2}{\pi}\frac{\Gamma}{\Gamma_0}\frac{q_d q}{h^2}\left(\frac{\rp}{r_0}\right)^{3/2-\alpha}r_0\Omega_0\nonumber\\
& &\left(1-\frac{4q_d\bar\xs}{q}\left(1-\frac{w_c}{w(\rp)}\right)\left(\frac{\rp}{r_0}\right)^{2-\alpha}\right)^{-1},
\end{eqnarray}
where 
\begin{equation}
m_c=\frac{4q_d\bar\xs}{q}\approx\frac{4q_d}{\sqrt{qh}},
\label{eqmc}
\end{equation}
where $\bar\xs = \xs/\rp$ and in the last step we have used $\bar\xs \approx\sqrt{q/h}$ \citep{horse}. 

\begin{figure}
\centering
\resizebox{\hsize}{!}{\includegraphics[]{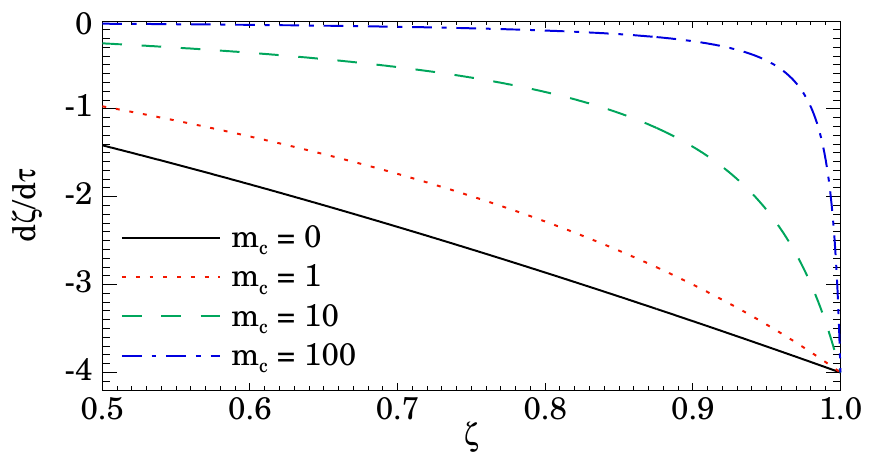}}
\caption{Migration rate versus radius for dynamic torques in an inviscid disc with $\alpha=0$ and $h=0.05$ for different values of $m_c$. The planet mass $q=10^{-5}$.}
\label{figInvisc}
\end{figure}

\subsubsection{Inviscid discs}
In the absence of viscosity, vortensity is conserved and we can take $w_c=w(r_0)$. Defining $\tau = \Omega_0t/\tau_\mathrm{mig}$ and $\zeta = \rp/r_0$, we get:
\begin{equation}
\left(\zeta^{\alpha-\frac{3}{2}}-m_c\left(1-\zeta^{\alpha-\frac{3}{2}}\right)\zeta^{\frac{1}{2}}\right)\frac{d\zeta}{d\tau}=\gamma,
\label{eqX}
\end{equation}
where $\gamma=\Gamma/\Gamma_0$, with $\Gamma$ the static torque, given by $\Gamma_L$ in an inviscid disc. Therefore, we expect $\gamma < 0$. The importance of the dynamical corotation torque is governed by a dimensionless parameter $m_c$, see equation (\ref{eqmc}), which is a measure of the mass inside the coorbital region in terms of the planet mass. For $m_c\ll 1$, dynamical corotation torques are unimportant, while for $m_c \gg 1$, they can significantly slow down inward migration if $\gamma<0$. For an Earth-mass planet located at 10 AU in the minimum mass Solar nebula with $h=0.05$, we find $m_c\approx 20$. The term involving $m_c$ in equation (\ref{eqX}) can be thought of as similar to the coorbital `mass' \footnote{For low-mass planets, it is a \emph{vortensity} deficit rather than a mass deficit.} deficit in classical Type III migration \citep{masset03}. Note that this deficit is built up through migration: it is zero at $\zeta=1$.

Equation (\ref{eqX}) can be integrated to
\begin{equation}
\gamma\tau = -\frac{1-\zeta^{\alpha-1/2}}{\alpha-1/2}+\frac{2m_c}{3}\left(1-\zeta^{3/2}\right)-\frac{m_c}{\alpha}\left(1-\zeta^\alpha\right),
\label{eqTau}
\end{equation}
taking $\zeta=1$ at $\tau=0$. For $\alpha=1/2$ the first on the right hand side should be interpreted as the limit $\alpha\rightarrow 1/2$, which equals $-\log \zeta$. The same goes for the last term on the right hand side for $\alpha=0$. It is clear that inward migration time scales get modified by roughly a factor $(1+m_c)$ for $\alpha < 3/2$. 

In general, equation (\ref{eqTau}) has to be solved numerically to obtain $\zeta(\tau)$. While this is straightforward, it is more convenient to work with migration rates $d\zeta/d\tau$ and compare these to numerical simulations in section \ref{secNum} below. Migration rates predicted by equation (\ref{eqX}) are depicted in figure \ref{figInvisc} for different values of $m_c$. Note that $m_c=0$ corresponds to migration due to static torques only, and that inward migration can be slowed down significantly for $m_c \gg 1$.
  
\subsubsection{Viscous discs}
\label{secVisc}
For discs with finite viscosity, we can no longer take $w_c=w(r_0)$. Viscous diffusion will act to take $w_c$ towards $w(\rp)$, on a time scale $\tau_\nu$:
\begin{equation}
\tau_\nu = \frac{\xs^2}{\nu} \approx \frac{q}{h}\frac{r_0^2\Omega_0}{\nu_0}\left(\frac{\rp}{r_0}\right)^{5/2-\alpha}\Omega_0^{-1}.
\end{equation}
The value $w(\rp)$ changes as the planet migrates, on a time scale of roughly $\tau_\mathrm{mig}$. 

A simple model can be constructed to incorporate viscosity. Consider the evolution of $w$, analogous to the saturation analysis in \cite{paardekooper11}. Neglecting vortensity advection in the horseshoe region for simplicity, the vortensity evolves, in the frame migrating with the planet, due to viscous diffusion and the migration of the planet. If the viscosity is low enough, vortensity is conserved and if the planet migrates up a gradient in vortensity, this means the difference in $w$ between the horseshoe region and the local disc increases. The latter can be incorporated by a source term inside the horseshoe region, changing the vortensity relative to the background as the planet migrates. Equation (10) from \cite{paardekooper11} then transforms to:
\begin{eqnarray}
\frac{\partial w(\bar x,t)}{\partial t}=\nonumber\\
\frac{3\nu_\mathrm{p}}{\rp^2}\left(1+\bar x\right)^{\frac{1}{2}}\frac{\partial}{\partial \bar x}\left(\left(1+\bar x\right)^{\frac{1}{2}}\frac{\partial}{\partial \bar x} \left(\left(1+\bar x\right)^{\alpha-\frac{3}{2}}w(\bar x,t)\right)\right) -\nonumber\\
\frac{d\rp}{dt}\left(\frac{3}{2}-\alpha\right)w_p \left(1+\bar x\right)^{\frac{1}{2}-\alpha}\Pi\left(\frac{\bar x}{\bar x_\mathrm{s}}\right),
\end{eqnarray}
where $\bar x = (r-\rp)/\rp$ and $\Pi$ denotes the rectangular function. The first term on the right-hand side accounts for viscous diffusion, while the second term is a source term, localised to the horseshoe region, changing the vortensity relative to the background as the planet migrates. Following \cite{paardekooper11}, we introduce $z = 2(1+\bar x)^{1/2}-2$ and $f=(1+z/2)^{2\alpha-3}w(\bar x,t)$ and look for stationary solutions:
\begin{equation}
\frac{d^2f}{dz^2}-\left(\frac{3}{2}-\alpha\right)\frac{\rp f_p}{3\nu_p}\frac{d\rp}{dt}\left(1+\frac{z}{2}\right)^{1-2\alpha}\Pi\left(\frac{z}{z_\mathrm{s}}\right)=0.
\end{equation}
Note that a stationary solution requires a constant migration rate. Since the migration rate changes on a time scale $\tau_\mathrm{mig}$, we need $\tau_\nu < \tau_\mathrm{mig}$, or
\begin{equation}
\frac{\nu_0}{r_0^2\Omega_0} > \frac{2}{\pi} \frac{q_d q^2}{h^3}.
\end{equation}
For smaller values of $\nu$, no stationary profile will develop, and the migration rate will to some extent be determined by the migration history (i.e. where the planet started), just as in the inviscid case.

Outside the horseshoe region, for $|z|>z_\mathrm{s}$, we have $f=f_p$. Expanding the second term for $z\ll1$ up to second order, and demanding that $f(z_\mathrm{s})=f(-z_\mathrm{s})=f_p$, we find that
\begin{equation}
\frac{f(z)}{f_p}=1+\left(\frac{3}{2}-\alpha\right)\frac{\rp}{3\nu_p}\frac{d\rp}{dt}\frac{z^2-z_\mathrm{s}^2}{2}.
\end{equation}
Taking the value at $z=0$ (i.e. $r=\rp$) as $w_c$, we get
\begin{equation}
1-\frac{w_c}{w(\rp)}=1-\frac{f(0)}{f_p}=\left(\frac{3}{2}-\alpha\right)\frac{\xs^2}{6\rp\nu_p}\frac{d\rp}{dt}.
\label{eqWc}
\end{equation}
As expected, the vortensity deficit is governed by the ratio of the viscous diffusion time scale across the horseshoe region and the migration time scale. The dynamical corotation torque is then given by
\begin{equation}
\Gamma_\mathrm{hs} = 2\pi\left(\frac{3}{2}-\alpha\right)\frac{\xs^3}{6\nu_p}\sigp\rp \op\left(\frac{d\rp}{dt}\right)^2,
\end{equation}  
leading to a second-order equation for the migration rate, with relevant solution
\begin{equation}
\frac{d\rp}{dt}=\Theta(k)\frac{2\gamma}{\pi}\frac{q_d q}{h^2} \left(\frac{\rp}{r_0}\right)^{3/2-\alpha} r_0\Omega_0,
\label{eqvVisc}
\end{equation}
with
\begin{equation}
k=\frac{8}{3\pi}\left(\frac{3}{2}-\alpha\right)\frac{\gamma q_d^2\bar\xs^3}{h^2}\frac{r_0^2\Omega_0}{\nu_0} \left(\frac{\rp}{r_0}\right)^{5-3\alpha},
\label{eqK}
\end{equation}
and
\begin{equation}
\Theta(k)=\frac{1-\sqrt{1-2k}}{k}.
\end{equation}
Here, $\gamma=\Gamma/\Gamma_0$ is the static part of the torque, including any (partially saturated) corotation torques. Note that $k$ depends on $\rp$, both explicitly and through $\gamma$ (whenever saturation of corotation torques plays a role).

\begin{figure}
\centering
\resizebox{\hsize}{!}{\includegraphics[]{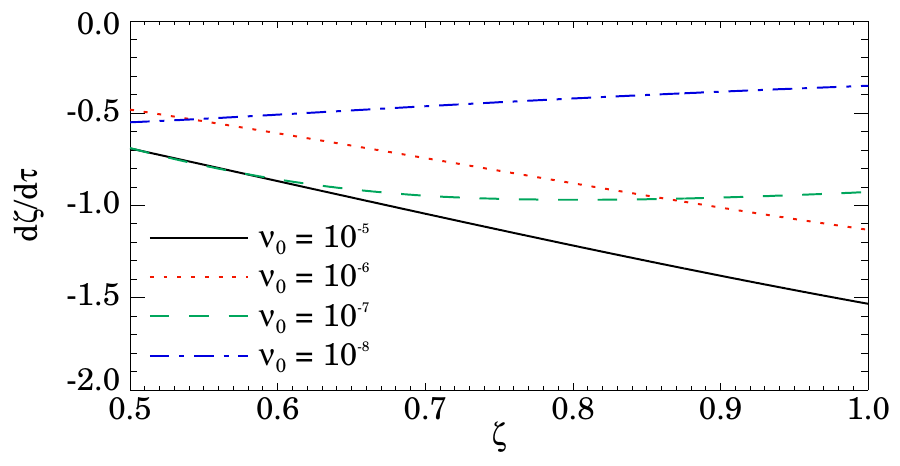}}
\caption{Migration rate versus radius for dynamic torques in a viscous disc with $\alpha=0$ and $h=0.05$ for different values of $\nu_0$. The planet mass is $q=10^{-5}$.}
\label{figVisc}
\end{figure}

A non-dimensional form of equation (\ref{eqvVisc}) reads
\begin{equation}
\frac{d\zeta}{d\tau}=\Theta(k)\gamma \zeta^{3/2-\alpha}.
\label{eqZeta}
\end{equation}
The effect of the dynamical corotation torque is contained in $\Theta$: for $|k| \ll 1$, $\Theta(k) \approx 1$ and migration proceeds according to static torques only. Note that $k<0$ if the static torque $\Gamma$ and the horseshoe drag differ in sign, and that $k\rightarrow 0$ for $\zeta\rightarrow 0$ for reasonable values of $\alpha$. Note also that $|k|\sim m_c\tau_\nu/\tau_\mathrm{mig}$. Examples are shown in figure \ref{figVisc}.

The importance of dynamical corotation torques can therefore be summarised as follows. If $m_c \ll 1$, migration will be due to static torques only. If $m_c > 1$ and $\tau_\nu < \tau_\mathrm{mig}$, dynamical torques will be important if $m_c\tau_\nu/\tau_\mathrm{mig} > 1$,  If $m_c > 1$ and $\tau_\nu > \tau_\mathrm{mig}$, dynamic torques will be important, but are more difficult to quantify. It is likely that the resulting migration will lie somewhere in between the viscous case and the inviscid case. Note that for $\tau_\nu \gg \tau_\mathrm{mig}$, the actual migration time scale will be $\sim m_c\tau_\mathrm{mig}$, which could make a stationary solution possible again. 

\subsection{Runaway migration}
Since $\Theta > 0$, the direction of migration is set by the static torque $\gamma$ (see equation (\ref{eqZeta})). If the dynamic corotation torque works against $\gamma$, $k<0$, and if $|k| \gg 1$, migration will slow down significantly, just as in the inviscid case. For $\alpha>3/2$, $k>0$ and $\Theta > 1$, and migration speeds up compared to the static case. If $k\rightarrow 1/2$, the assumption of steady migration breaks down, and there is a possibility of a runaway. Note that having $k\approx 1$ amounts to demanding that the coorbital `mass' deficit, as given by equation (\ref{eqWc}) multiplied by $m_c$, should be of order unity. This is the same criterion as for classical Type III migration: the mass deficit should be of the order of the planet mass \citep{masset03}. Noting that the saturation parameter $p^2\propto \zeta^{1-\alpha}$, we see that for $\alpha > 3/2$, inward migration leads to more saturated static corotation torques. Hence we expect $|\gamma|$ to decrease as the planet migrates, since the negative contribution of the corotation torque decreases in magnitude. For $\alpha < 5/3$, $k$ decreases as well as the planet migrates inward, making dynamic torques less important. Therefore, unless $k$ can be of order unity with very little migration, inward runaway migration requires a very steep surface density profile $\alpha > 5/3$. 

Since the static torque $\gamma$ in viscous discs also includes a contribution from the unsaturated corotation torque, there is a possibility that $\gamma>0$. In this case, we again have $k>0$ (since we need $\alpha < 3/2$ in order to have a positive corotation torque) and the possibility of outward runaway migration. For $\alpha < 1$, outward migration again leads to more saturated static corotation torques. In an isothermal disc, we need $\alpha < -1$ for $\gamma$ to become positive \citep{drag}. While this may be an unrealistic density profile, it can serve as an approximate model for non-isothermal discs with strong entropy-related corotation torques. Since the static corotation torque becomes more saturated as the planet migrates outward, at some point $\gamma$ will change sign and become negative. If at this point we have $k \ll 1$, migration will stall at this zero-torque radius. However, if we can reach $k\sim 1$ before this happens, runaway outward migration may take over, in a sense beating saturation and moving the planet beyond the zero-torque radius. An approximate criterion would be to have $k >1/2$ at the initial location of the planet, which can be translated into a condition on the viscosity:
\begin{equation}
\frac{\nu_0}{r_0^2\Omega_0} < \frac{16}{3\pi}\left(\frac{3}{2}-\alpha\right)\frac{\gamma q_d^2 \bar\xs^3}{h^2}.
\label{eqCritVisc}
\end{equation} 
As expected, runaway migration is more likely in massive discs. However, since $\bar\xs^2\propto q$, it is also more likely for more massive planets, until they become massive enough to open up gaps. Using equation (\ref{eqP}), we can cast above condition in terms of $p$:
\begin{equation}
p > \sqrt{\frac{\pi}{8\gamma}\frac{1}{\frac{3}{2}-\alpha}}\frac{h}{q_d}.
\end{equation}
If we want to rely on unsaturated static corotation torques to yield $\gamma>0$, we need $p\approx 1$, and therefore a disc that is a few times more massive than the MMSN for outward runaway migration to occur. These estimates will be tested using numerical hydrodynamical simulations in the next section. 

\section{Numerical results}
\label{secRes}

In this section, we compare the results of the previous section to numerical hydrodynamical simulations. We will first look at cases where the static torque is negative, in which case dynamic torques can act to slow down inward migration, and secondly discs where the static torque can be positive, leading in some cases to runaway outward migration.

\begin{figure}
\centering
\resizebox{\hsize}{!}{\includegraphics[]{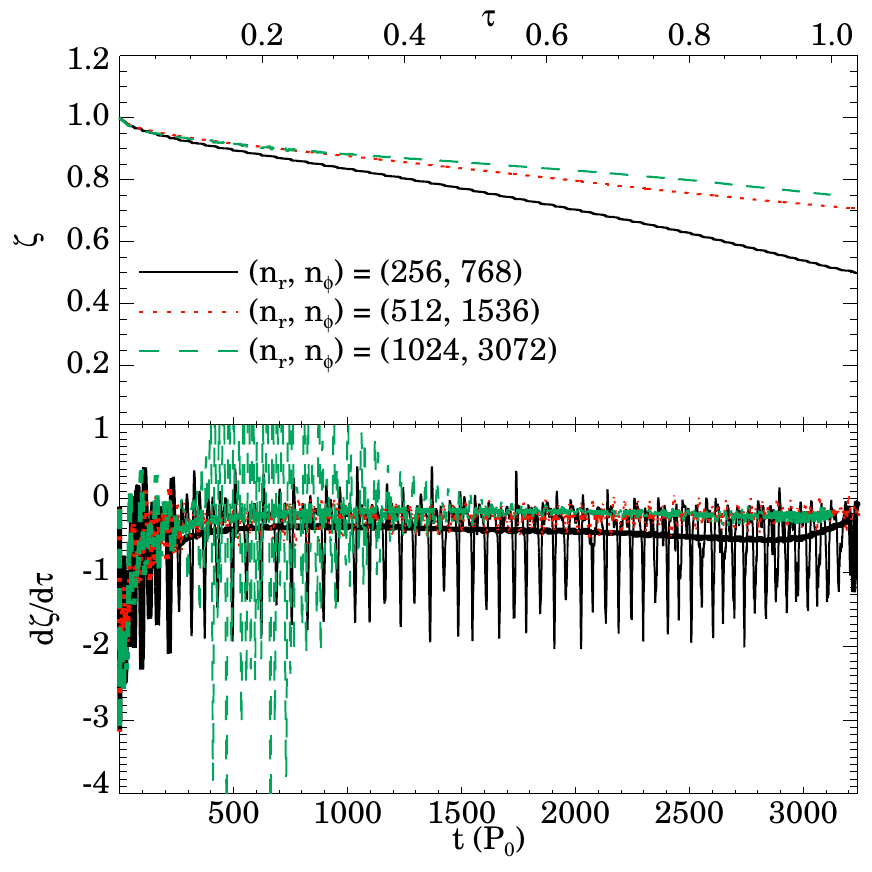}}
\caption{Numerical results for an inviscid disc at different resolutions. Top panel: dimensionless orbital radius. Bottom panel: dimensionless migration rate.}
\label{figNumInvisc}
\end{figure}

\subsection{Slowing down inward migration}
\label{secInward}

\begin{figure*}
\centering
\resizebox{\hsize}{!}{\includegraphics[]{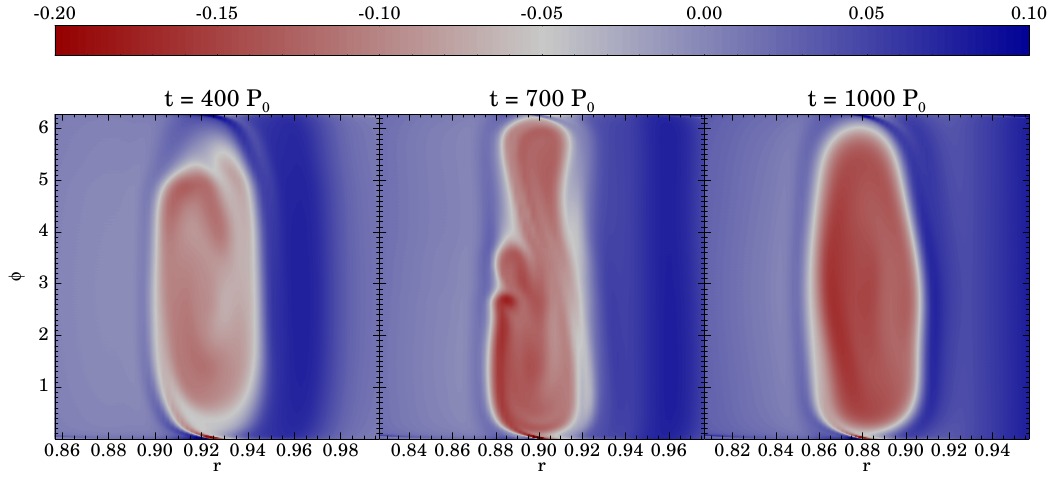}}
\caption{Vortensity perturbation near the radial location of the planet at three different times for an inviscid disc with $(n_r, n_\varphi)=(1024, 3072)$.}
\label{figVort}
\end{figure*}

In this section, we first focus on a disc with constant surface density ($\alpha=0$). This means the static torque is negative, driving inward migration, while the static corotation torque is positive. Therefore, the dynamical corotation torque will be positive and work against inward migration. Note that planets with $\xs \sim h$ experience a boost in the static corotation torque due to their larger-than-expected value of $\xs$ \citep{masset06,horse}, making the static torque in fact positive. However, this effect only kicks in at planet masses $q\sim 10^{-4}$ for our adopted value of $h=0.05$, while we focus on $q=10^{-5}$ in this section.  

We first look at an inviscid disc. According to equation (\ref{eqX}), migration will slow down considerably whenever $m_c$ is much larger than unity. In order to emphasise the effects of $m_c$, we choose a very massive disc with $q_d=0.02$, ten times more massive than the MMSN at 10 AU. This leaves us with $m_c\sim 100$, which should have a considerable effect on the migration rate, see figure \ref{figInvisc}.

In figure \ref{figNumInvisc}, we show the evolution of the semi-major axis of the planet, together with its migration rate, for three different resolutions. For the lowest resolution, the scale height is resolved by $\sim$ 6 cells, and $\xs$ by 2 cells. For the wave torque, it is important to resolve $H$, while $\xs$ needs to be resolved in order to get the corotation torque right. It seems unlikely that 2 grid cells will be enough to achieve this, and indeed from figure \ref{figNumInvisc} we see that the run with $(n_r,n_\varphi)=(256,768)$ shows a significantly faster migration rate compared to the higher resolution runs. This can be attributed to numerical diffusion of vortensity into the horseshoe region. In deriving equation (\ref{eqX}), we assumed $w_0=w(r_0)$. In the presence of numerical diffusion, this no longer holds. While numerical diffusion is unlikely to act as a real viscosity, it nevertheless will diminish the vortensity difference between the horseshoe region and the rest of the disc, akin to the viscous case studied in section \ref{secVisc}. The planet can not maintain its vortensity deficit, and dynamical torques are less efficient in slowing down migration. In completely inviscid discs, numerical diffusion will play a role at any resolution at late enough times. It can be seen that while the two highest resolution runs agree very well for the first 1500 orbits, at later times the run with the lower resolution speeds up in comparison, again because numerical diffusion reduces the vortensity deficit of the planet.

Another difference between the two highest resolution cases can be seen in the migration rates in the bottom panel of figure \ref{figNumInvisc}. The run with the highest resolution shows large amplitude variations between $t=400$ $P_0$ and $t=1500$ $P_0$. These variations are caused by vortices inside the corotation region, which form while the disc adjusts to the large vortensity gradients set up by migration, much like what happens due to the mixing of vortensity as the corotation torque saturates \citep{balmforth01}. Three snapshots of the vortensity perturbation are shown in figure \ref{figVort}. In the left panel, the vortensity perturbation is still relatively mild, but as the planet migrates further in, it gets larger until vortices appear at the edge of the corotation region (middle panel). In the right panel, the disc has almost finished adjusting to the new situation, leaving a laminar flow with a definite vortensity deficit. Note that in this simulation, $\xs$ is resolved by $\sim$ 8 grid cells, which is enough to resolve these vortices. The run with $(n_r, n_\varphi)=(512, 1536)$ has not enough resolution and these vortices do not appear. It is interesting that these vortices, while they cause large spikes in the torque and migration rate, do not affect the global migration rate. Indeed, the highest resolution run only starts to diverge \emph{after} the vortices have disappeared, see figure \ref{figNumInvisc}. This is linked to the fact that these vortices are located inside the corotation region, which means that their effect on the torque averages out over long time scales. 

\begin{figure}
\centering
\resizebox{\hsize}{!}{\includegraphics[]{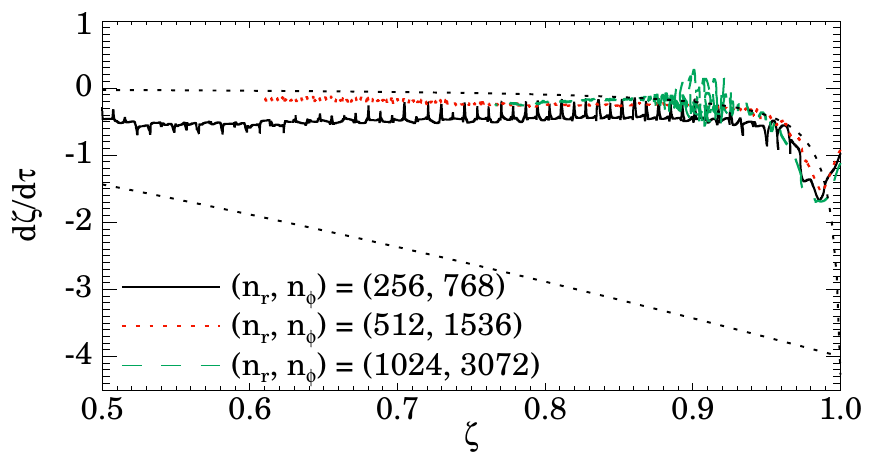}}
\caption{Dimensionless migration rate versus dimensionless orbital radius for an inviscid disc with $\alpha=0$ and $q_d=0.02$ at three different resolutions. The bottom dotted black curve is the prediction from static torques only, while the top dotted curve is the prediction of equation (\ref{eqX}).}
\label{figNumInviscAdot}
\end{figure}

We can compare the numerical migration rates to the prediction of equation (\ref{eqX}) by looking at $\zeta$ versus $d\zeta/d\tau$, see figure \ref{figNumInviscAdot}. The two black dotted curves indicate the predictions from static torques (bottom curve) and equation (\ref{eqX}) (top curve). It is clear that, for this massive disc, the orbital evolution of embedded planets is remarkably different from that predicted by static torques alone. For $\zeta > 0.85$, the highest resolution run falls nicely on the prediction of equation (\ref{eqX}). Note again the effect of the vortices around $\zeta=0.9$. For smaller $\zeta$, or, equivalently, at later times, numerical diffusion takes the migration rate to more negative values, depending on the resolution. It is likely that increasing the resolution even further will make the migration rate follow equation (\ref{eqX}) for longer times.

Results for different surface density profiles are very similar as long as $\alpha < 3/2$. For $\alpha=3/2$, corotation torques (static or dynamic) are small, and migration proceeds according to the static wave torque only. Steeper density profiles show an acceleration of inward migration due to dynamic torques, as expected. We come back to this case, where the dynamical torque is working in the direction of migration, in section \ref{secRunaway} below. Results for lower values of $m_c$ also show good agreement with equation (\ref{eqX}).

\begin{figure}
\centering
\resizebox{\hsize}{!}{\includegraphics[]{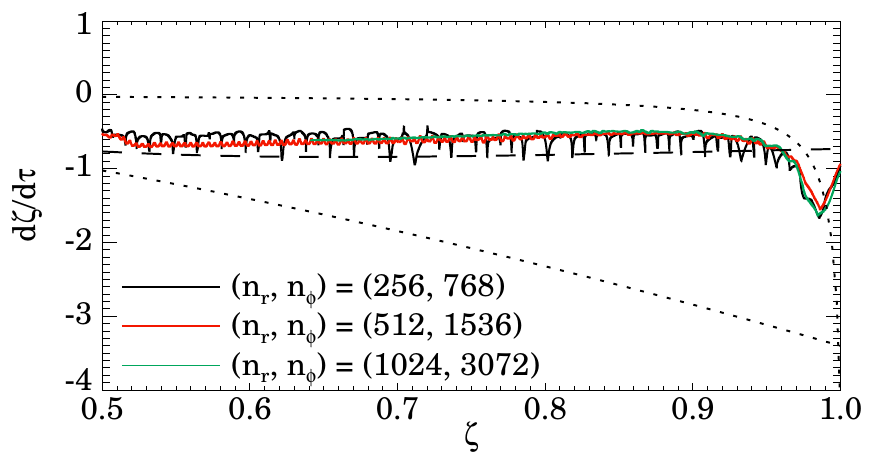}}
\caption{As figure \ref{figNumInviscAdot}, but for a viscous disc with $\nu_0=10^{-7} r_0^2\Omega_0$. The black dashed curve denotes the prediction from equation (\ref{eqZeta}).}
\label{figNumViscAdot}
\end{figure}

We now turn our attention to viscous discs, for which our simple model prediction is given by equation (\ref{eqZeta}). We still take $\alpha=0$, so that the static corotation torque is positive, but now take $\nu_0 = 10^{-7} r_0^2\Omega_0$, which makes $k\sim 10$. For this level of viscosity, corotation torques will be saturated to a large extent ($p^2 \approx 32$). The resulting migration rates are displayed in figure \ref{figNumViscAdot}, for the same three resolutions as in figure \ref{figNumInviscAdot}. Note that the static torques (lower dotted curve) are slightly different from figure \ref{figNumInviscAdot} because of partly unsaturated static corotation torques. The prediction of equation (\ref{eqX}) (upper dotted curve) does not change since it does not depend on viscosity. Equation (\ref{eqZeta}) is represented by the dashed curve. The first thing to note about the numerical results is that all resolutions agree to a very large extent, contrary to the inviscid case. Therefore, we can conclude that numerical diffusion for the lowest resolution is comparable to or lower than a viscosity of $10^{-7}$. The numerical migration rates agree with the simple model to approximately $20\%$. Around $\zeta=0.5$, the planet comes too close to the inner boundary (located at $\zeta=0.4$), which affects the migration rate. The initial drop in migration rate close to $\zeta=1$ is caused by the finite time needed to build up a vortensity deficit. Note that this is not part of the viscous model, which deals with the equilibrium state only.

\begin{figure}
\centering
\resizebox{\hsize}{!}{\includegraphics[]{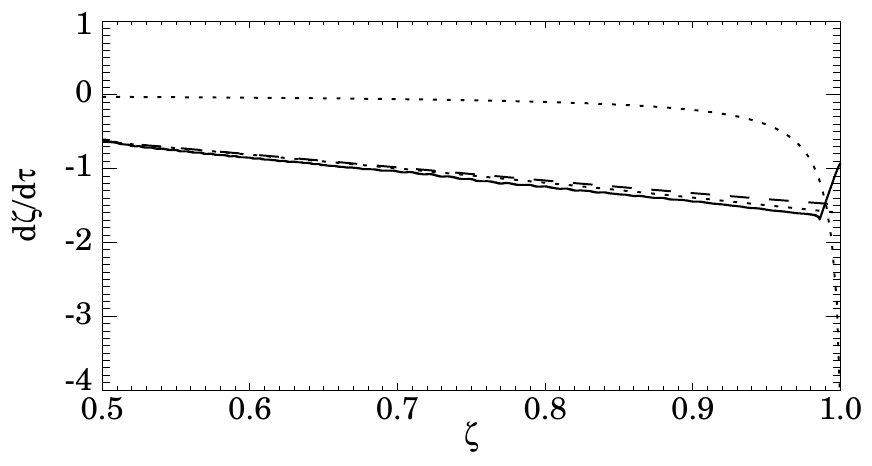}}
\caption{Dimensionless migration rate versus dimensionless orbital radius for a disc with $\alpha=0$, $q_d=0.02$ and $\nu_0=10^{-5} r_0^2\Omega_0$, at resolution $(n_r, n_\varphi)=(512, 1536)$. The bottom dotted black curve is the prediction from static torques only, while the top dotted curve is the prediction of equation (\ref{eqX}).}
\label{figNumViscAdot2}
\end{figure}

Increasing the level of viscosity reduces the vortensity deficit, as expected. Results for $\nu_0=10^{-5} r_0^2\Omega_0$ are shown in figure \ref{figNumViscAdot2}. For this value of $\nu_0$, we find $k\sim 0.1$, which makes $\Theta\approx 1.05$. Equation (\ref{eqZeta}) therefore predicts no significant departure from static torque migration, and this is confirmed by the numerical results in figure \ref{figNumViscAdot2}. Note that compared to figure \ref{figNumViscAdot}, the static torque is less negative because viscosity keeps the corotation torque from saturating. 

\subsection{Runaway outward migration}
\label{secRunaway}

Now we turn to a disc in which outward migration is possible. In realistic cases, this requires a non-isothermal setup \citep{paard06}.  However, in order to study its possible effects under the current approximation we can boost the corotation torque by choosing a surface density profile that increases outwards \citep{drag}. While admittedly unrealistic, it can serve as a proof of principle of what can happen in non-isothermal discs if similar dynamical corotation torques can be set up. Moreover, it can shed light on whether static and dynamic torques can act together to overcome the strong wave torque. 

\begin{figure}
\centering
\resizebox{\hsize}{!}{\includegraphics[]{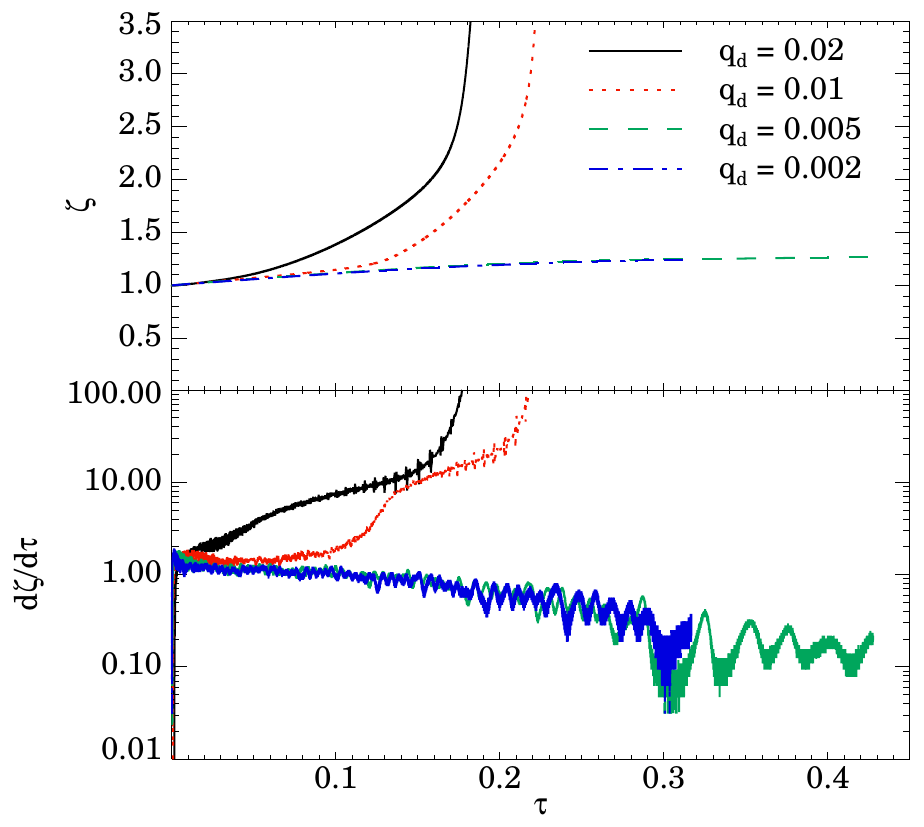}}
\caption{Numerical results for a disc with $h=0.05$, $\nu_0=10^{-6}r_0^2\Omega_0$ and $\alpha=-2$, with a planet mass $q=10^{-5}$ for different disc masses. Top panel: dimensionless orbital radius. Bottom panel: dimensionless migration rate. Note that time has been scaled with $\tau_\mathrm{mig}$.}
\label{figOutward}
\end{figure}

We start by looking at a disc with $h=0.05$ and $\alpha=-2$, put in our standard planet with $q=10^{-5}$ and vary the disc mass. The viscosity $\nu_0=10^{-6} r_0^2\Omega_0$ is chosen such that at $\zeta=1$, corotation torques are able to overcome the wave torque, resulting in outward Type I migration. However, since $p^2\propto \zeta^{1-\alpha}$ (see equations (\ref{eqP}) and (\ref{eqnu})), corotation torques will become more and more saturated as the planet migrates outward. Migration due to static torques would come to a halt around $\zeta=1.3$. Numerical results are presented in figure \ref{figOutward}. The two largest disc masses clearly lead to runaway outward migration, overcoming saturation because dynamic torques take over. The planets reach the outer boundary of the grid, located at $\zeta=4.6$, before they stop. The simulations with the two lowest disc masses do not show evidence for runaway migration. 

\begin{figure}
\centering
\resizebox{\hsize}{!}{\includegraphics[]{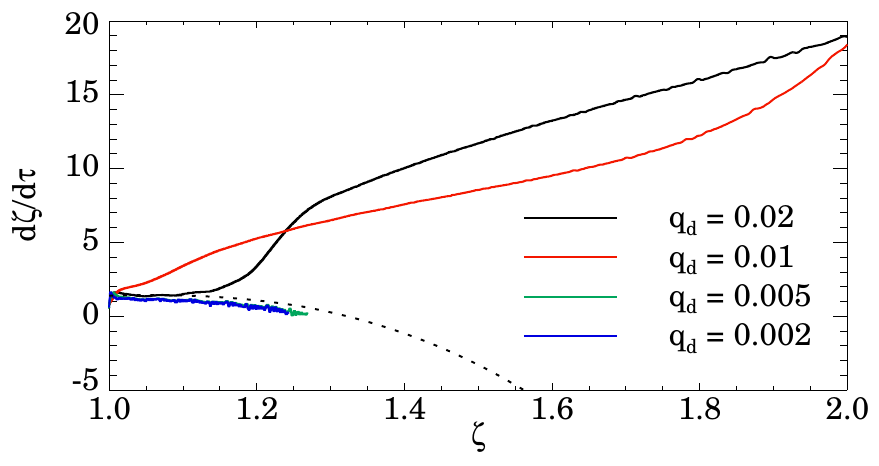}}
\caption{Dimensionless migration rate versus dimensionless orbital radius for a disc with $\alpha=-2$, $h=0.05$ and $\nu_0=10^{-6} r_0^2\Omega_0$ for different disc masses. The dotted black curve is the prediction from static torques only.}
\label{figOutwardAdot}
\end{figure}

Figure \ref{figOutwardAdot} shows the migration rate versus radius for the same four simulations as figure \ref{figOutward}, together with the prediction of static torque migration (dotted curve). As mentioned before, static torques predict that migration should come to a halt around $\zeta=1.3$. The results for the two lowest disc masses closely follow the prediction of the static torques. The values of the parameter $k$ for these four simulations are, from highest to lowest disc mass, $2.3$, $0.58$, $0.15$, $0.023$. The results of figures \ref{figOutward} and \ref{figOutwardAdot} are therefore consistent with the prediction that runaway migration can occur if $k > 1/2$. 
 
\begin{figure}
\centering
\resizebox{\hsize}{!}{\includegraphics[]{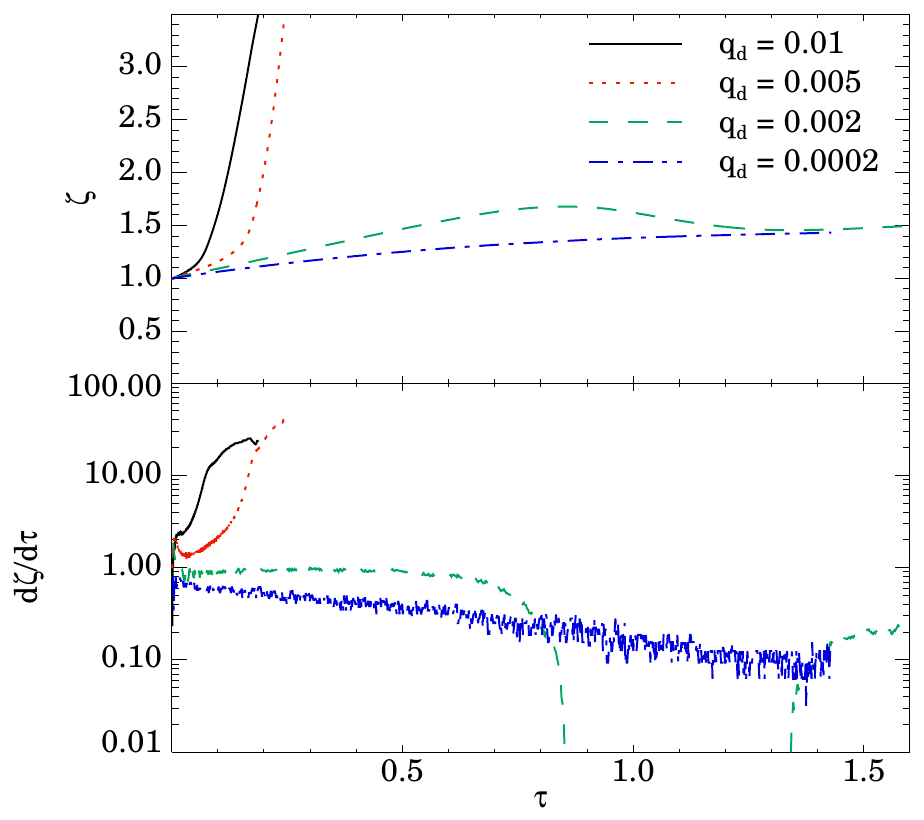}}
\caption{Numerical results for a disc with $h=0.05$, $\nu_0=10^{-6}r_0^2\Omega_0$ and $\alpha=0$, with a planet mass $q=10^{-4}$ for different disc masses. Top panel: dimensionless orbital radius. Bottom panel: dimensionless migration rate. Note that time has been scaled with $\tau_\mathrm{mig}$.}
\label{figOutwardHigh}
\end{figure}

While we are considering low-mass planets in this study, for which $q < h^3$ and therefore $\bar\xs<h$ \citep{horse}, it is known that planets approaching this limit experience an enhanced corotation torque \citep{masset06}. This is because they have a wider horseshoe region than predicted for low-mass planets \citep{horse}. Not only are corotation torques potentially very strong for these planets, their wider horseshoe regions, and the resulting short libration time scales, make saturation more of a problem. These intermediate-mass planets are therefore interesting candidates to study the effects of dynamical torques. 

The results for $q=10^{-4}$ ($31$ $M_\oplus$ around a Solar mass star) in a disc with constant surface density ($\alpha=0$) are depicted in figure \ref{figOutwardHigh}. For this planet, $q/h^3=0.8$, so it only just qualifies as a low-mass planet. The enhanced corotation torque allows us to study outward migration for a less extreme surface density profile, while we increase the viscosity to $\nu_0=2\cdot 10^{-5} r_0^2\Omega_0$ to keep the corotation torque strong enough at $\zeta=1$ and to make sure no significant surface density depression can form near the orbit of the planet. As in the case with $q=10^{-5}$, static torques predict that outward migration will come to a halt, this time around $\zeta=1.5$. 

It is clear that dynamical torques play an even stronger role for these intermediate mass planets. Not only can we get away with $\alpha=0$ to induce outward migration, runaway migration sets in at lower disc masses. Even the MMSN at 10 AU, $q_d=0.002$, appears to be borderline, with an episode of relatively fast outward migration. This fast migration is not sustained, however, and is followed by a period of inward migration, after which the planet sets off again to larger values of $\zeta$. Reducing the disc mass by a factor of 10 results in migration under influence of static torques only. For $q_d=0.002$, we find $k \sim 0.1$, and therefore runaway migration should not occur according to the requirements of the simple model presented in section \ref{secMig}. Note however, that $k$ will increase steeply with $\zeta$ according to equation (\ref{eqK}), even for $\alpha=0$, so that conditions for runaway migration may improve as the planet moves outward. On the other hand, viscosity changes with radius, which affects the saturation of the static corotation torque. The $k$-criterion should therefore be taken only as a rough indication of the possibility of runaway migration.  

\section{Discussion}
\label{secDisc}

\subsection{Comparison with previous work}

In order to see dynamical corotation torques on low-mass planets in action in numerical simulations, four criteria have to be fulfilled: the planet needs to be free to change its orbit, the viscosity in the disc has to be low enough (the critical level depends on disc mass, see below), the horseshoe region should be adequately resolved, and the equilibrium disc should allow for static corotation torques. To our knowledge, no study so far has fulfilled all of these, and hence dynamical corotation torques on low-mass planets have not been observed so far.  
  
Work on low-mass planet migration has for a large part been focused on measuring static torques on planets kept on fixed orbits \citep[e.g.][]{dangelo02, dangelo03, bate03, masset06, drag, masset09, paardekooper10}. Freely evolving planets were studied by \cite{baruteau08}, who showed that one either has to include self-gravity of the disc, or remove the axisymmetric component of the force on the planet in order to obtain consistent results. While \cite{baruteau08} study freely migrating low-mass planets, their simulations were probably not run for long enough to see the effects of dynamical torques at large disc mass. Longer term simulations of Type I migration in self-gravitating discs at low viscosities were presented in \cite{li09} and \cite{yu10}. However, they focused on a strictly isothermal disc with $\alpha=3/2$, for which there is no corotation torque (static or dynamic). Other studies involving free low-mass planets include for example \cite{cresswell07}, who performed 3D simulations of inclined and eccentric planets. While they were focused on eccentricity and inclination damping, their disc mass was small ($q_d \approx 10^{-3}$), and because they were doing 3D simulations their resolution was comparatively low (roughly half our lowest resolution of $(n_r, n_\varphi)=(256, 768)$). It is therefore unlikely that they could have seen any dynamical torques in action. 

\begin{figure}
\centering
\resizebox{\hsize}{!}{\includegraphics[]{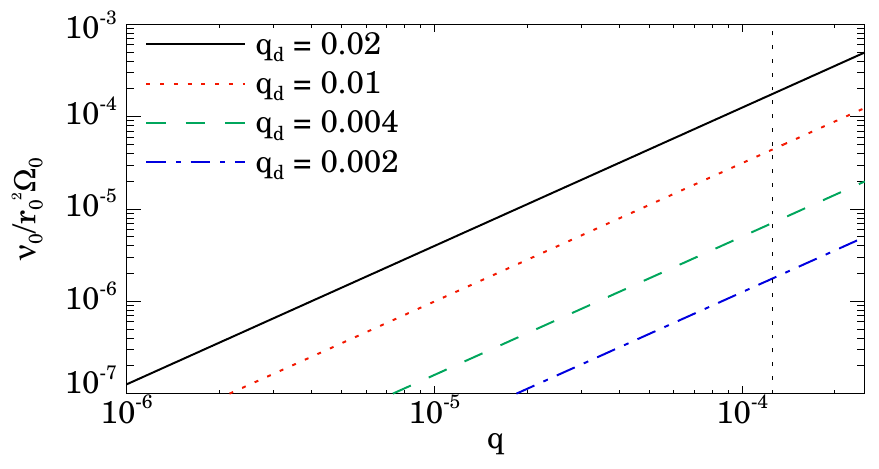}}
\caption{Critical viscosity below which dynamical corotation torques play a role, as determined by equation (\ref{eqCritVisc}), for $\alpha=0$ and different disc masses. The vertical dotted line denotes $q=h^3$, above which the approximation of small surface density perturbations breaks down.}
\label{figCritVisc}
\end{figure}

In general, most studies take a standard value of viscosity to be $\nu_0 \sim 10^{-5}$, for which dynamic torques on low-mass planets are relatively unimportant (e.g. figure \ref{figNumViscAdot2}). The critical viscosity below which dynamical corotation torques can be important is essentially given by equation (\ref{eqCritVisc}). This critical viscosity is displayed in figure \ref{figCritVisc} for different disc and planet masses. These curves were constructed assuming $\bar\xs = \sqrt{q/h}$ and may therefore underestimate the critical viscosity towards higher planet masses. We can only apply the analysis presented in this paper to the left of the vertical dotted line, where $q < h^3$. The MMSN at 10 AU, represented by the dash-dotted line ($q_d=0.002$), needs $\nu_0 < 10^{-6} r_0^2\Omega_0$ for dynamical torques to play a role for any planet mass. It gets more interesting for more massive discs: twice the MMSN brings dynamical torques into the picture for $\sim 30$ $M_\oplus$ planets ($q=10^{-4}$) at standard levels of viscosity. 

\cite{ogilvie06} studied the effect of planetary migration on the strength of a single corotation resonance. They showed that a vortensity anomaly can be set up for fluid elements trapped in the corotation region, which modifies the corotation torque. While \cite{ogilvie06} were concerned mainly with the effect of corotation torques on the eccentricity evolution of the planet, they speculated that similar effects could play a role in the more difficult problem of the coorbital region of a planet. Our results show that this is indeed the case.

\subsection{Comparison with classical Type III migration}

While dynamical corotation torques can be understood within the framework of Type III migration, there are a few important differences between classical Type III migration as presented in \cite{masset03}. In that work, a coorbital mass deficit was created by the planet because it carved a partial gap. This deficit has a positive feedback on migration. Low-mass planets, on the other hand, do not perturb the surface density distribution of the disc significantly, but still a `mass' deficit can arise if there is a radial gradient in vortensity in the disc. This is because in Type III migration, the coorbital mass gets weighted by the vortensity \citep{masset03}. While for high-mass planets, a real \emph{mass} deficit can arise, for low-mass planets it is rather a \emph{vortensity} deficit (obtained through migration), but the effect is the same.

Another difference between classical Type III migration and dynamical corotation torques presented here, is that, since dynamical corotation torques on low-mass planets are driven directly by background gradients in the unperturbed disc, they can be directed inward or outward. The feedback on migration can therefore be either positive or negative. Negative feedback was presented in section \ref{secInward}, positive feedback in section \ref{secRunaway}. In classical Type III migration, the coorbital mass deficit is always positive, since the planet carves a partial gap, resulting in a positive feedback on migration. 

The condition for runaway migration to occur is similar to classical Type III migration: the coorbital `mass' deficit should be comparable to the planet mass. The way this deficit is obtained is different in the case of low-mass planets, though: rather than carving a partial gap, these planets can create a `mass' deficit by migrating into a region of different vortensity.
 
\subsection{Simplifying assumptions}

In order to keep the problem tractable, physically and numerically, several simplifications were introduced. The possible impact of these assumptions are discussed briefly below. 

While dynamical torques can slow down inward migration considerably in massive discs, the most dramatic effects occur when corotation torques and migration work in the same direction. In reality, a non-isothermal disc is needed to get corotation torques to overpower the wave torque. In this paper, we have focused on the simple, (locally) isothermal case, leaving a study of more realistic discs to future work. However, a few comments can be made here. 

In section \ref{secRunaway}, we have mimicked strong positive corotation torques as might arise in non-isothermal discs by choosing a surface density profile that increases outward. It should be born in mind that this choice has other effects as well, besides boosting the corotation torque. Looking at equation (\ref{eqK}), we see that for $\alpha=-2$, $k$ increases with $\zeta$ as $\zeta^{11}$. This means that as the planet migrates outward, it gets easier to start runaway migration. Note that we expect this for any $\alpha<3/2$, but $\alpha=-2$ leads to quite a dramatic change in $k$ with radius, and perhaps to an overly optimistic view of the onset of runaway outward migration. A second comment that should be made here is that it is not immediately obvious that the non-isothermal version of the horseshoe drag can give rise to dynamical torques in the same way as in the isothermal case. The non-isothermal horseshoe drag is dominated by an edge term at the separatrix, where sheets of vortensity are created because of a discontinuity in the entropy profile \citep{masset09, paardekooper10}. While migration in nearly adiabatic discs can lead to a strong difference between the entropy inside the corotation region and that of the unperturbed disc, and could therefore boost the corotation torque, it is not clear if such a situation can be readily incorporated in the framework presented here. 

An alternative approach to study runaway migration on longer time scales in more realistic isothermal discs would be to enforce outward migration for a few libration time scales, which leads to a vortensity deficit, before releasing the planet \citep[this approach was also taken in][]{masset03}. It would be interesting to see how far outward the planet can migrate with, say, $\alpha=1/2$. However, one drawback of this approach is that it is likely that the results will depend on how large the initial vortensity deficit is, which in turn depends on the speed (and, in inviscid discs, the distance) of the forced outward migration. It is difficult to come up with a realistic value for this parameter, which in the end must be determined by non-isothermal simulations.

Viscosity plays a major role in determining the strength of dynamical torques. It should be kept in mind that a Navier-Stokes viscosity is used to model the effects of angular momentum transport due to turbulence. Discs can be turbulent for example in regions where the magneto-rotational instability (MRI) operates \citep{balbus91}. Unfortunately, global MRI simulations for thousands of orbits, as needed for a study of dynamical torques, are beyond current computational capabilities. It is however encouraging that MRI simulations show reasonably good agreement with viscous disc simulations as far as static corotation torques are concerned \citep{baruteau11}. 

If planets migrate radially, then not only does the local level of viscosity play a role, the viscosity profile becomes important as well. We have taken the simple approach of choosing $\nu=\nu(r)$ so that the initial surface density profile is an equilibrium solution. In section \ref{secRunaway}, for example, the viscosity profile was such that saturation of the static torque increased as the planet moved outward. In reality, this does not have to be the case. The level of turbulence in the disc as a function of radius strongly depends on for example the ionisation fraction in the disc; for an overview of recent developments in this area see \cite{turner14}.

We have neglected accretion onto the planet. While this is reasonable for Earth-like protoplanets, gas accretion may play a role for the larger mass planet ($q=10^{-4}$) we have considered. It was found, for example, that runaway outward migration of massive planets (comparable to Jupiter) is usually stalled because of accretion \citep{peplinski08c}. However, in our case, runaway outward migration is actually easier for higher-mass planets (up to $q \sim h^3$), so that accretion onto the planet may in fact play a favourable role. On the other hand, accretion onto the planet will change the local surface density distribution, and it is not clear how this will affect the torques onto the planet.

We have ignored the self-gravity of the disc. Note that the Toomre stability parameter $Q=h/q_d$ is $2.5$ at $\zeta=1$ for our largest disc mass. Including self-gravity would lead to an enhanced wave torque \citep{pierens05, baruteau08}, and perhaps at our highest disc mass, depending on the local cooling time, gravito-turbulence \citep[e.g.][]{gammie01}. Migration in such turbulent discs was studied for example in \cite{baruteau11b}. 
   
The final important simplification made in this work is the 2D approximation. Fortunately, 2D and 3D simulations usually give qualitatively similar results, with a tendency for 3D simulations to show stronger effects of the corotation torque \citep{kley09}. Unfortunately, 3D simulations at adequate resolution (resolving the horseshoe region for low-mass planets) for thousands of orbits is computationally still very difficult. Note that a nested or adaptive grid is of limited use: the whole horseshoe region, spanning $2\pi$ in azimuth, has to be resolved. 

\subsection{Inertial mass}

It should be noted that dynamical corotation torques occur for planets that do not open up significant surface density depressions. While there can be strong perturbations in vortensity, as long as $\xs \ll H$ these do not translate into surface density perturbations \citep{casoli09}. Even though at sufficiently low viscosities, even low-mass planets may open gaps \citep{goodman01, rafikov02}, dynamical torques can be seen to operate in viscous discs as well, as long as the disc mass is high enough. In this sense, the mechanism of dynamical torques should be distinguished from the ``inertial mass" as defined by \cite{hourigan84} and \cite{ward89}. The concept of this inertial mass also relies on a feedback of migration on the torque, but this time on the wave torque. Wave damping through viscosity or shocks modifies the density profile around the planet, which has a negative effect on the migration rate \citep[see also][]{rafikov02}. In contrast, dynamical corotation torques can either provide positive or negative feedback on the migration rate, depending on the background vortensity gradient. The inertial mass was studied numerically in \cite{li09} and \cite{yu10} for an isothermal disc with $\alpha=3/2$, for which corotation torques (static and dynamic) should be absent. 

\subsection{Consequences for planet formation}

Dynamical torques are important in discs more massive than the MMSN. They can significantly slow down inward migration, but perhaps the most interesting effects occur when the planet is migrating in the direction set by static corotation torques, which is outward unless the surface density falls off very rapidly with radius ($\alpha > 3/2$). While these static torques may saturate at a particular location in the disc, to an extent that migration stalls, if the disc is massive enough dynamic corotation torques will take over, possibly taking the planet into runaway outward migration. The planet can therefore migrate across this zero-torque line set by saturation. The limit of this outward migration is difficult to predict. In our simple disc models, conditions for runaway outward migration improve as the planet migrates outward, and as a consequence the planet could migrate all the way to the outer grid boundary. In reality, conditions in the disc can change drastically over such a large distance (a factor of 4 in radius): the level of viscosity, background gradients in temperature and surface density, and, in the non-isothermal case, the cooling time. It is very well possible that as the planet migrates outward, it will lose its vortensity deficit after which Type III outward migration will come to a halt. Such processes are best studied in non-isothermal discs, that do not require artificial surface density profiles to induce outward migration. 

\section{Conclusions}
\label{secCon}

We have presented an analysis of torques on migrating, low-mass planets in locally isothermal discs. In addition to static torques, that do not depend on the migration rate, planets experience dynamical torques whenever there is a radial gradient in vortensity. These dynamical torques, proportional to the migration rate, can have either a positive or a negative feedback on migration, depending on whether the planet is migrating ``with" or ``against" the static corotation torque. In discs a few times more massive than the MMSN, if the viscosity is low enough, the effects of dynamical torques can be profound: inward migration can be slowed down significantly, and outward migration can proceed beyond zero-torque lines set by saturation in a runaway fashion. The latter case is especially interesting if it can occur in non-isothermal discs, since it can increase the region in the disc where outward migration is possible.

\section*{Acknowledgements}
SJP is supported by a Royal Society University Research Fellowship. This work was performed using the Darwin Supercomputer of the University of Cambridge High Performance Computing Service (http://www.hpc.cam.ac.uk), provided by Dell Inc. using Strategic Research Infrastructure Funding from the Higher Education Funding Council for England.

\bibliography{paardekooper.bib}

\begin{thebibliography}{}

\bibitem[\protect\citeauthoryear{{Balbus} \& {Hawley}}{{Balbus} \&
  {Hawley}}{1991}]{balbus91}
{Balbus} S.~A.,  {Hawley} J.~F.,  1991, \apj, 376, 214

\bibitem[\protect\citeauthoryear{{Balmforth} \& {Korycansky}}{{Balmforth} \&
  {Korycansky}}{2001}]{balmforth01}
{Balmforth} N.~J.,  {Korycansky} D.~G.,  2001, \mnras, 326, 833

\bibitem[\protect\citeauthoryear{{Baruteau}, {Crida}, {Paardekooper}, {Masset},
  {Guilet}, {Bitsch}, {Nelson}, {Kley} \& {Papaloizou}}{{Baruteau}
  et~al.}{2013}]{baruteau13}
{Baruteau} C.,  {Crida} A.,  {Paardekooper} S.-J.,  {Masset} F.,  {Guilet} J.,
  {Bitsch} B.,  {Nelson} R.~P.,  {Kley} W.,    {Papaloizou} J.~C.~B.,  2013,
  ArXiv e-prints

\bibitem[\protect\citeauthoryear{{Baruteau}, {Fromang}, {Nelson} \&
  {Masset}}{{Baruteau} et~al.}{2011}]{baruteau11}
{Baruteau} C.,  {Fromang} S.,  {Nelson} R.~P.,    {Masset} F.,  2011, \aap,
  533, A84

\bibitem[\protect\citeauthoryear{{Baruteau} \& {Masset}}{{Baruteau} \&
  {Masset}}{2008}]{baruteau08}
{Baruteau} C.,  {Masset} F.,  2008, \apj, 678, 483

\bibitem[\protect\citeauthoryear{{Baruteau}, {Meru} \&
  {Paardekooper}}{{Baruteau} et~al.}{2011}]{baruteau11b}
{Baruteau} C.,  {Meru} F.,    {Paardekooper} S.-J.,  2011, \mnras, 416, 1971

\bibitem[\protect\citeauthoryear{{Bate}, {Lubow}, {Ogilvie} \& {Miller}}{{Bate}
  et~al.}{2003}]{bate03}
{Bate} M.~R.,  {Lubow} S.~H.,  {Ogilvie} G.~I.,    {Miller} K.~A.,  2003,
  \mnras, 341, 213

\bibitem[\protect\citeauthoryear{{Bitsch}, {Crida}, {Morbidelli}, {Kley} \&
  {Dobbs-Dixon}}{{Bitsch} et~al.}{2013}]{bitsch13}
{Bitsch} B.,  {Crida} A.,  {Morbidelli} A.,  {Kley} W.,    {Dobbs-Dixon} I.,
  2013, \aap, 549, A124

\bibitem[\protect\citeauthoryear{{Bitsch} \& {Kley}}{{Bitsch} \&
  {Kley}}{2011}]{bitsch11}
{Bitsch} B.,  {Kley} W.,  2011, \aap, 536, A77

\bibitem[\protect\citeauthoryear{{Casoli} \& {Masset}}{{Casoli} \&
  {Masset}}{2009}]{casoli09}
{Casoli} J.,  {Masset} F.~S.,  2009, \apj, 703, 845

\bibitem[\protect\citeauthoryear{{Cresswell}, {Dirksen}, {Kley} \&
  {Nelson}}{{Cresswell} et~al.}{2007}]{cresswell07}
{Cresswell} P.,  {Dirksen} G.,  {Kley} W.,    {Nelson} R.~P.,  2007, \aap, 473,
  329

\bibitem[\protect\citeauthoryear{{Crida} \& {Morbidelli}}{{Crida} \&
  {Morbidelli}}{2007}]{crida07}
{Crida} A.,  {Morbidelli} A.,  2007, \mnras, 377, 1324

\bibitem[\protect\citeauthoryear{{D'Angelo}, {Henning} \& {Kley}}{{D'Angelo}
  et~al.}{2002}]{dangelo02}
{D'Angelo} G.,  {Henning} T.,    {Kley} W.,  2002, \aap, 385, 647

\bibitem[\protect\citeauthoryear{{D'Angelo}, {Kley} \& {Henning}}{{D'Angelo}
  et~al.}{2003}]{dangelo03}
{D'Angelo} G.,  {Kley} W.,    {Henning} T.,  2003, \apj, 586, 540

\bibitem[\protect\citeauthoryear{{Gammie}}{{Gammie}}{2001}]{gammie01}
{Gammie} C.~F.,  2001, \apj, 553, 174

\bibitem[\protect\citeauthoryear{{Goldreich} \& {Tremaine}}{{Goldreich} \&
  {Tremaine}}{1980}]{goldreich80}
{Goldreich} P.,  {Tremaine} S.,  1980, \apj, 241, 425

\bibitem[\protect\citeauthoryear{{Goodman} \& {Rafikov}}{{Goodman} \&
  {Rafikov}}{2001}]{goodman01}
{Goodman} J.,  {Rafikov} R.~R.,  2001, \apj, 552, 793

\bibitem[\protect\citeauthoryear{{Hayashi}}{{Hayashi}}{1981}]{hayashi81}
{Hayashi} C.,  1981, Progress of Theoretical Physics Supplement, 70, 35

\bibitem[\protect\citeauthoryear{{Hourigan} \& {Ward}}{{Hourigan} \&
  {Ward}}{1984}]{hourigan84}
{Hourigan} K.,  {Ward} W.~R.,  1984, \icarus, 60, 29

\bibitem[\protect\citeauthoryear{{Kley}, {Bitsch} \& {Klahr}}{{Kley}
  et~al.}{2009}]{kley09}
{Kley} W.,  {Bitsch} B.,    {Klahr} H.,  2009, \aap, 506, 971

\bibitem[\protect\citeauthoryear{{Korycansky} \& {Papaloizou}}{{Korycansky} \&
  {Papaloizou}}{1996}]{kory96}
{Korycansky} D.~G.,  {Papaloizou} J.~C.~B.,  1996, \apjs, 105, 181

\bibitem[\protect\citeauthoryear{{Li}, {Lubow}, {Li} \& {Lin}}{{Li}
  et~al.}{2009}]{li09}
{Li} H.,  {Lubow} S.~H.,  {Li} S.,    {Lin} D.~N.~C.,  2009, \apjl, 690, L52

\bibitem[\protect\citeauthoryear{{Lin} \& {Papaloizou}}{{Lin} \&
  {Papaloizou}}{1986a}]{lin86a}
{Lin} D.~N.~C.,  {Papaloizou} J.,  1986a, \apj, 307, 395

\bibitem[\protect\citeauthoryear{{Lin} \& {Papaloizou}}{{Lin} \&
  {Papaloizou}}{1986b}]{lin86b}
{Lin} D.~N.~C.,  {Papaloizou} J.,  1986b, \apj, 309, 846

\bibitem[\protect\citeauthoryear{{Masset}}{{Masset}}{2000}]{masset00}
{Masset} F.,  2000, \aaps, 141, 165

\bibitem[\protect\citeauthoryear{{Masset}}{{Masset}}{2001}]{masset01}
{Masset} F.~S.,  2001, \apj, 558, 453

\bibitem[\protect\citeauthoryear{{Masset}}{{Masset}}{2008}]{masset08}
{Masset} F.~S.,  2008, in {Sun} Y.-S.,  {Ferraz-Mello} S.,   {Zhou} J.-L.,
  eds, IAU Symposium Vol. 249, IAU Symposium. p.~331

\bibitem[\protect\citeauthoryear{{Masset} \& {Casoli}}{{Masset} \&
  {Casoli}}{2009}]{masset09}
{Masset} F.~S.,  {Casoli} J.,  2009, \apj, 703, 857

\bibitem[\protect\citeauthoryear{{Masset} \& {Casoli}}{{Masset} \&
  {Casoli}}{2010}]{masset10}
{Masset} F.~S.,  {Casoli} J.,  2010, \apj, 723, 1393

\bibitem[\protect\citeauthoryear{{Masset}, {D'Angelo} \& {Kley}}{{Masset}
  et~al.}{2006}]{masset06}
{Masset} F.~S.,  {D'Angelo} G.,    {Kley} W.,  2006, \apj, 652, 730

\bibitem[\protect\citeauthoryear{{Masset} \& {Papaloizou}}{{Masset} \&
  {Papaloizou}}{2003}]{masset03}
{Masset} F.~S.,  {Papaloizou} J.~C.~B.,  2003, \apj, 588, 494

\bibitem[\protect\citeauthoryear{{Nelson}, {Papaloizou}, {Masset} \&
  {Kley}}{{Nelson} et~al.}{2000}]{nelson00}
{Nelson} R.~P.,  {Papaloizou} J.~C.~B.,  {Masset} F.,    {Kley} W.,  2000,
  \mnras, 318, 18

\bibitem[\protect\citeauthoryear{{Ogilvie} \& {Lubow}}{{Ogilvie} \&
  {Lubow}}{2006}]{ogilvie06}
{Ogilvie} G.~I.,  {Lubow} S.~H.,  2006, \mnras, 370, 784

\bibitem[\protect\citeauthoryear{{Paardekooper}, {Baruteau}, {Crida} \&
  {Kley}}{{Paardekooper} et~al.}{2010}]{paardekooper10}
{Paardekooper} S.-J.,  {Baruteau} C.,  {Crida} A.,    {Kley} W.,  2010, \mnras,
  401, 1950

\bibitem[\protect\citeauthoryear{{Paardekooper}, {Baruteau} \&
  {Kley}}{{Paardekooper} et~al.}{2011}]{paardekooper11}
{Paardekooper} S.-J.,  {Baruteau} C.,    {Kley} W.,  2011, \mnras, 410, 293

\bibitem[\protect\citeauthoryear{{Paardekooper} \& {Mellema}}{{Paardekooper} \&
  {Mellema}}{2006}]{paard06}
{Paardekooper} S.-J.,  {Mellema} G.,  2006, \aap, 459, L17

\bibitem[\protect\citeauthoryear{{Paardekooper} \& {Papaloizou}}{{Paardekooper}
  \& {Papaloizou}}{2009a}]{drag}
{Paardekooper} S.-J.,  {Papaloizou} J.~C.~B.,  2009a, \mnras, 394, 2283

\bibitem[\protect\citeauthoryear{{Paardekooper} \& {Papaloizou}}{{Paardekooper}
  \& {Papaloizou}}{2009b}]{horse}
{Paardekooper} S.-J.,  {Papaloizou} J.~C.~B.,  2009b, \mnras, 394, 2297

\bibitem[\protect\citeauthoryear{{Pepli{\'n}ski}, {Artymowicz} \&
  {Mellema}}{{Pepli{\'n}ski} et~al.}{2008a}]{peplinski08}
{Pepli{\'n}ski} A.,  {Artymowicz} P.,    {Mellema} G.,  2008a, \mnras, 386, 164

\bibitem[\protect\citeauthoryear{{Pepli{\'n}ski}, {Artymowicz} \&
  {Mellema}}{{Pepli{\'n}ski} et~al.}{2008b}]{peplinski08c}
{Pepli{\'n}ski} A.,  {Artymowicz} P.,    {Mellema} G.,  2008b, \mnras, 387,
  1063

\bibitem[\protect\citeauthoryear{{Pierens} \& {Hur{\'e}}}{{Pierens} \&
  {Hur{\'e}}}{2005}]{pierens05}
{Pierens} A.,  {Hur{\'e}} J.-M.,  2005, \aap, 433, L37

\bibitem[\protect\citeauthoryear{{Rafikov}}{{Rafikov}}{2002}]{rafikov02}
{Rafikov} R.~R.,  2002, \apj, 572, 566

\bibitem[\protect\citeauthoryear{{Syer} \& {Clarke}}{{Syer} \&
  {Clarke}}{1995}]{syer95}
{Syer} D.,  {Clarke} C.~J.,  1995, \mnras, 277, 758

\bibitem[\protect\citeauthoryear{{Tanaka}, {Takeuchi} \& {Ward}}{{Tanaka}
  et~al.}{2002}]{tanaka02}
{Tanaka} H.,  {Takeuchi} T.,    {Ward} W.~R.,  2002, \apj, 565, 1257

\bibitem[\protect\citeauthoryear{{Turner}, {Fromang}, {Gammie}, {Klahr},
  {Lesur}, {Wardle} \& {Bai}}{{Turner} et~al.}{2014}]{turner14}
{Turner} N.~J.,  {Fromang} S.,  {Gammie} C.,  {Klahr} H.,  {Lesur} G.,
  {Wardle} M.,    {Bai} X.-N.,  2014, ArXiv e-prints

\bibitem[\protect\citeauthoryear{{Walsh}, {Morbidelli}, {Raymond}, {O'Brien} \&
  {Mandell}}{{Walsh} et~al.}{2011}]{walsh11}
{Walsh} K.~J.,  {Morbidelli} A.,  {Raymond} S.~N.,  {O'Brien} D.~P.,
  {Mandell} A.~M.,  2011, \nat, 475, 206

\bibitem[\protect\citeauthoryear{{Ward}}{{Ward}}{1991}]{ward91}
{Ward} W.~R.,  1991, in Lunar and Planetary Institute Conference Abstracts.
  p.~1463

\bibitem[\protect\citeauthoryear{{Ward} \& {Hourigan}}{{Ward} \&
  {Hourigan}}{1989}]{ward89}
{Ward} W.~R.,  {Hourigan} K.,  1989, \apj, 347, 490

\bibitem[\protect\citeauthoryear{{Yu}, {Li}, {Li}, {Lubow} \& {Lin}}{{Yu}
  et~al.}{2010}]{yu10}
{Yu} C.,  {Li} H.,  {Li} S.,  {Lubow} S.~H.,    {Lin} D.~N.~C.,  2010, \apj,
  712, 198

\end{thebibliography}

\label{lastpage}
\end{document}